\documentclass[final,5p,times,twocolumn]{elsarticle}

\usepackage{graphicx}

\usepackage{upgreek} 
\input myunits.sty
\input mymath.sty

\usepackage{amssymb}
\usepackage{amsthm}

\usepackage{lineno}

\journal{Nuclear Instruments and Methods A}

\begin{document}

\begin{frontmatter}



\title{A Time Projection Chamber for High-Rate Experiments: Towards an
Upgrade of the ALICE TPC}


\author{Bernhard Ketzer}
\author{for the GEM-TPC and ALICE TPC Collaborations}
\address{Technische Universit\"at M\"unchen, Physik Department,
  85748 Garching, Germany}

\begin{abstract}
  A Time Projection Chamber (TPC) is a powerful detector for
  3-dimensional tracking and particle identification for ultra-high
  multiplicity events. It is the central tracking device of many
  experiments, e.g.\ the ALICE experiment at CERN. 
  The
  necessity of a switching electrostatic gate, which prevents ions
  produced  
  in the amplification region of MWPCs from entering the drift volume,
  however,   
  restricts its application to trigger rates of the order of $1\,\kHz$. 

  Charge amplification by Gas Electron Multiplier (GEM) foils instead of
  proportional wires 
  offers an intrinsic suppression of the ion backflow, although not to
  the same level as a gating grid.  
  Detailed Monte Carlo simulations have shown that the distortions due
  to residual space charge from back-drifting ions can be limited
  to a few $\Cm$, and thus 
  can be corrected using standard calibration techniques.  
  A prototype GEM-TPC has been built 
  with the largest active volume to date for a detector of this type. 
  It
  has been 
  commissioned with 
  cosmics and particle beams at the FOPI experiment at GSI, and was
  employed for a physics measurement with pion beams. 

  For future operation of the ALICE TPC at the CERN LHC beyond 2019,
  where Pb-Pb collision rates of $50\,\kHz$ are expected, 
  it is planned to replace the existing MWPCs by GEM detectors,
  operated in a continuous, triggerless readout mode, thus allowing an
  increase in 
  event rate by a factor of 100. 
  As a first step of the R\&D program, a prototype of an Inner Readout
  Chamber was equipped with large-size GEM foils and exposed to
  beams of protons, pions and electrons from the CERN PS. 
  
  In this paper, 
  new results are shown 
  concerning ion backflow, spatial and momentum resolution of the
  FOPI GEM-TPC, detector 
  calibration, and $\diff{E}/\diff{x}$ resolution with both detector
  prototypes. The 
  perspectives of a GEM-TPC for ALICE with continuous 
  readout will be discussed.
\end{abstract}


\end{frontmatter}



\section{Introduction}
A Time Projection Chamber (TPC) \cite{Nygren:1978rx} is a
high-resolution 
detector 
providing three-dimensional tracking of charged particles even in very
high multiplicity environments such as heavy ion collisions, e.g.\ at  
the ALICE experiment at CERN's LHC \cite{Alme:2010ke}.
The large number of space points measured along each particle track
considerably eases the task of pattern recognition for complex events,
and in addition allows the identification of each particle by measuring its
specific energy loss. 
Since track
reconstruction requires a precise knowledge of the electric and
magnetic fields inside the chamber, distortions have to be kept at a
minimum. To this end, TPCs are normally used in a gated mode, where an
electrostatic grid blocks the ions created in the amplification stage,
usually an array of proportional wires, before they reach the drift
volume. This introduces inevitable dead times and limits the maximum
trigger rate to a few kHz. 
After the second long shutdown of the LHC in 2018, a luminosity for
Pb-Pb collisions of $6\EE{27}\,\Cm^{-2}\,\s^{-1}$ is expected.
In order to make full use of the increase in luminosity, ALICE plans 
to record all minimum bias events at a 
rate of about 
$50\,\kHz$ \cite{Musa:1475243}, about two orders of magnitude
higher than at present. 
As a result particle tracks from 5 events on average will be superimposed 
in the drift volume. 
A continuous, untriggered readout of
the TPC is the obvious mode of operation, 
precluding the use of a gating grid. 

\section{A GEM-based TPC}
The use of Gas Electron Multiplier foils \cite{Sauli:97} instead of
proportional wires 
offers an intrinsic suppression of the ion backflow
\cite{Sauli:2003yf}, although not to 
the level of $<10^{-4}$ as with a gating grid. Suppression factors around
or below the percent 
level have been reached \cite{Blatt:2006nx}, but strongly depend on the gas
composition and the magnetic field. 
Together with low-noise readout electronics, 
allowing an operation at gas gains of the order of $10^{3}$, this
results in about $10$ ions drifting back into the drift volume per
electron arriving at the amplification stage. 
The drift  
distortions caused by the ion space charge have been shown in
simulations to be of the order of 
a few $\Cm$, which can be corrected using standard calibration
techniques \cite{Bohmer:2012wd,Musa:1475243}, provided the ion charge
distribution is stable over time scales of the order of $\ms$.   

\begin{figure}[tbp] 
  \centering 
  \includegraphics[width=0.45\textwidth,keepaspectratio]{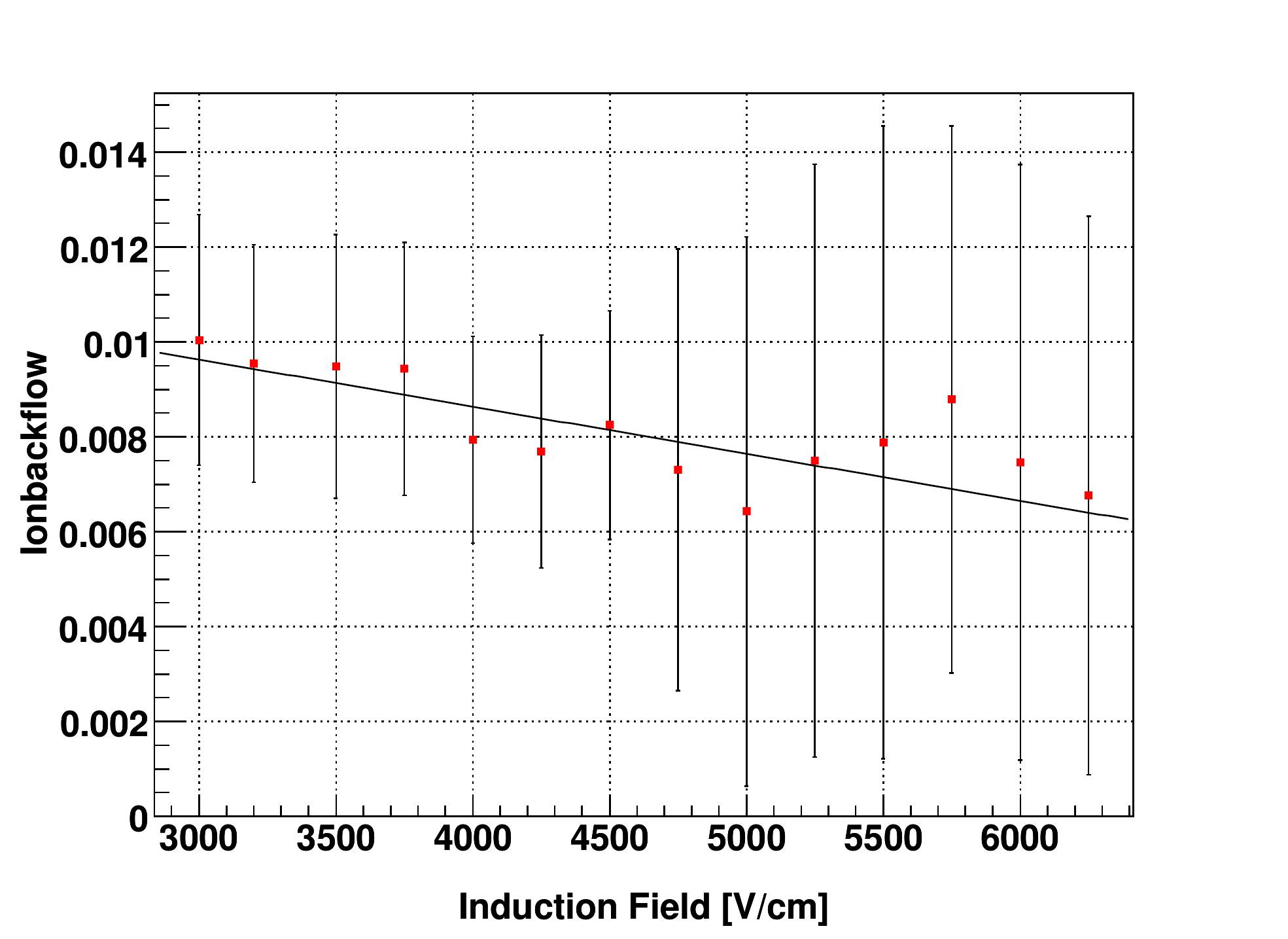}
  \caption{Ion backflow, defined as the ratio of cathode to anode
    currents, in an Ar/CO$_2$ (70/30) mixture, as a function of
    induction field. The other fields are set to
    $E_\mathrm{drift}=250\,\V/\Cm$, $E_\mathrm{trans1}=6.0\,\kV/\Cm$,
    $E_\mathrm{trans2}=160\,\V/\Cm$, the potentials across the GEMs
    are $\Delta U_1=330\,\V$, $\Delta U_2=375\,\V$, $\Delta
    U_1=450\,\V$, resulting in a gain of about $10^{4}$.}
  \label{fig:IB-vs-Eind_ar-70_co2-30}
\end{figure}
\begin{figure}[tbp] 
  \centering 
  \includegraphics[width=0.45\textwidth,keepaspectratio]{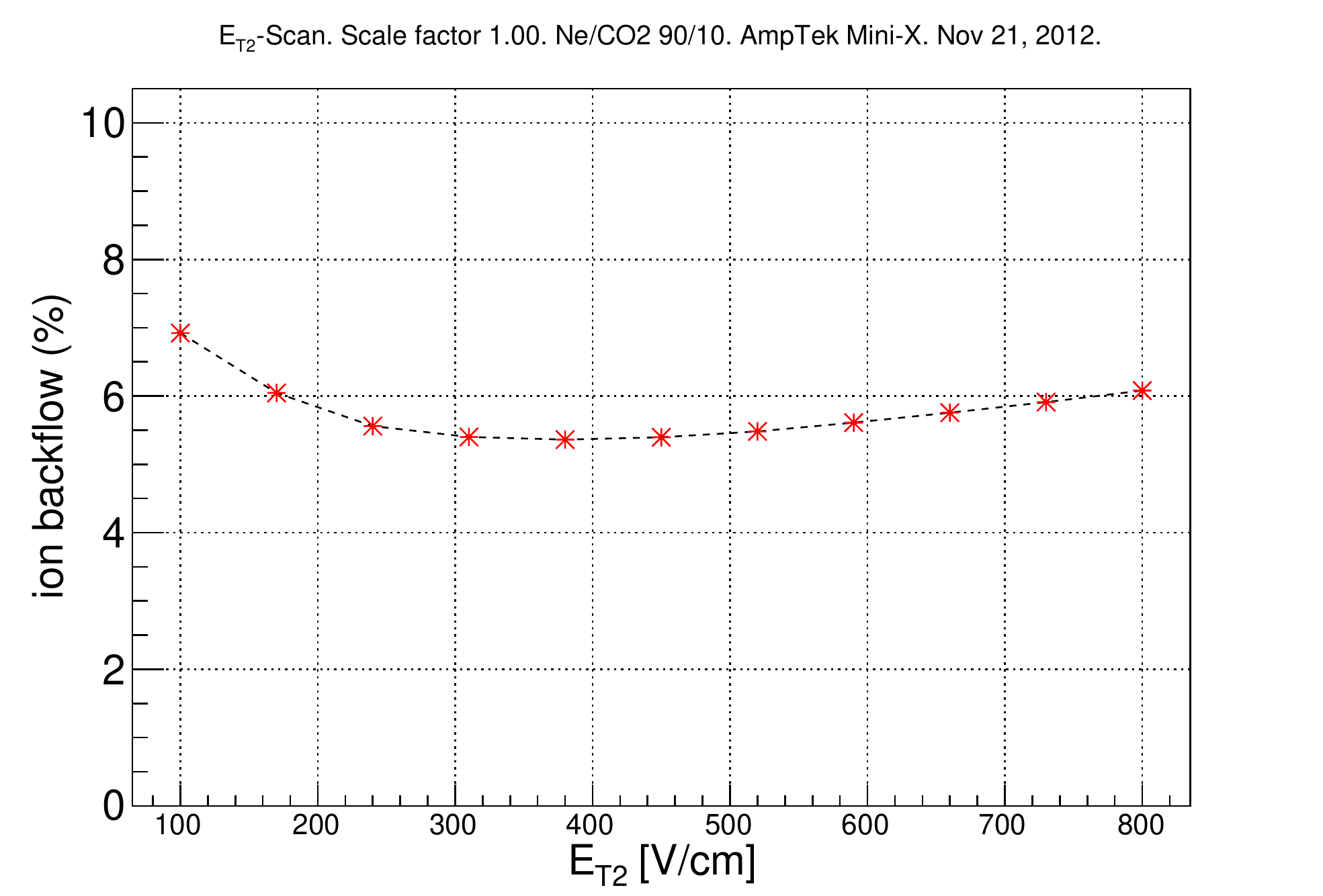}
  \caption{Ion backflow in a Ne/CO$_2$ (90/10) mixture, as a function of
    transfer field 2. The other fields are set to
    $E_\mathrm{drift}=400\,\V/\Cm$, $E_\mathrm{trans1}=3.8\,\kV/\Cm$,
    $E_\mathrm{ind}=3.8\,\kV/\Cm$, the potentials across the GEMs
    are $\Delta U_1=225\,\V$, $\Delta U_2=235\,\V$, $\Delta
    U_1=285\,\V$, resulting in a gain of about $10^{3}$.}
  \label{fig:IB-vs-Et2_ne-90_co2-10}
\end{figure}

In order to minimize distortions for the case of the ALICE TPC, which
operates in a low solenoid field of $0.5\,\T$, 
measurements of the crucial ion backflow ($I\!B$), defined as the ratio of
cathode to anode current,  
have been performed in triple GEM detectors with Ar- and Ne-based gas
mixtures without magnetic
field. Figure~\ref{fig:IB-vs-Eind_ar-70_co2-30} shows this quantity 
for an Ar/CO$_2$ (70/30) mixture 
as a function of the induction field, i.e.\ the field between the last
GEM and the readout anode. Values of the $I\!B$ of about $0.8\%$ have
been reached in a configuration with a low drift field, alternatingly
high and low 
electric fields between GEMs, and the highest gain in the last GEM.  
It is to be noted that such a configuration is not optimal for
stability of the chamber against discharges in case of highly ionizing
particles \cite{Bachmann:01e}. Moving to a Ne/CO$_2$ (90/10) mixture
as currently used in the ALICE TPC requires a higher drift
field ($400\,\V/\Cm$) and lower fields between GEMs ($<4\,\kV/\Cm$ in
order not to 
enter a regime of 
charge amplification). As a consequence, the $I\!B$ is higher by about a
factor of 5, as can be
seen in Fig.~\ref{fig:IB-vs-Et2_ne-90_co2-10}. Steps to further
decrease this value include adding small admixtures of N$_2$ to the
gas and a fourth GEM, guided by microscopic simulations of charge
transfer processes in GEMs \cite{GARFIELDPP:2012}.  
Strategies to eliminate the problem completely, e.g.\ by using foils
with additional patterning on one side \cite{Lyashenko:09a} are under
investigation, but are currently limited by the size of foils.   

\section{The FOPI GEM-TPC}
\begin{figure}[tbp] 
  \centering 
  \includegraphics[width=0.45\textwidth,keepaspectratio]{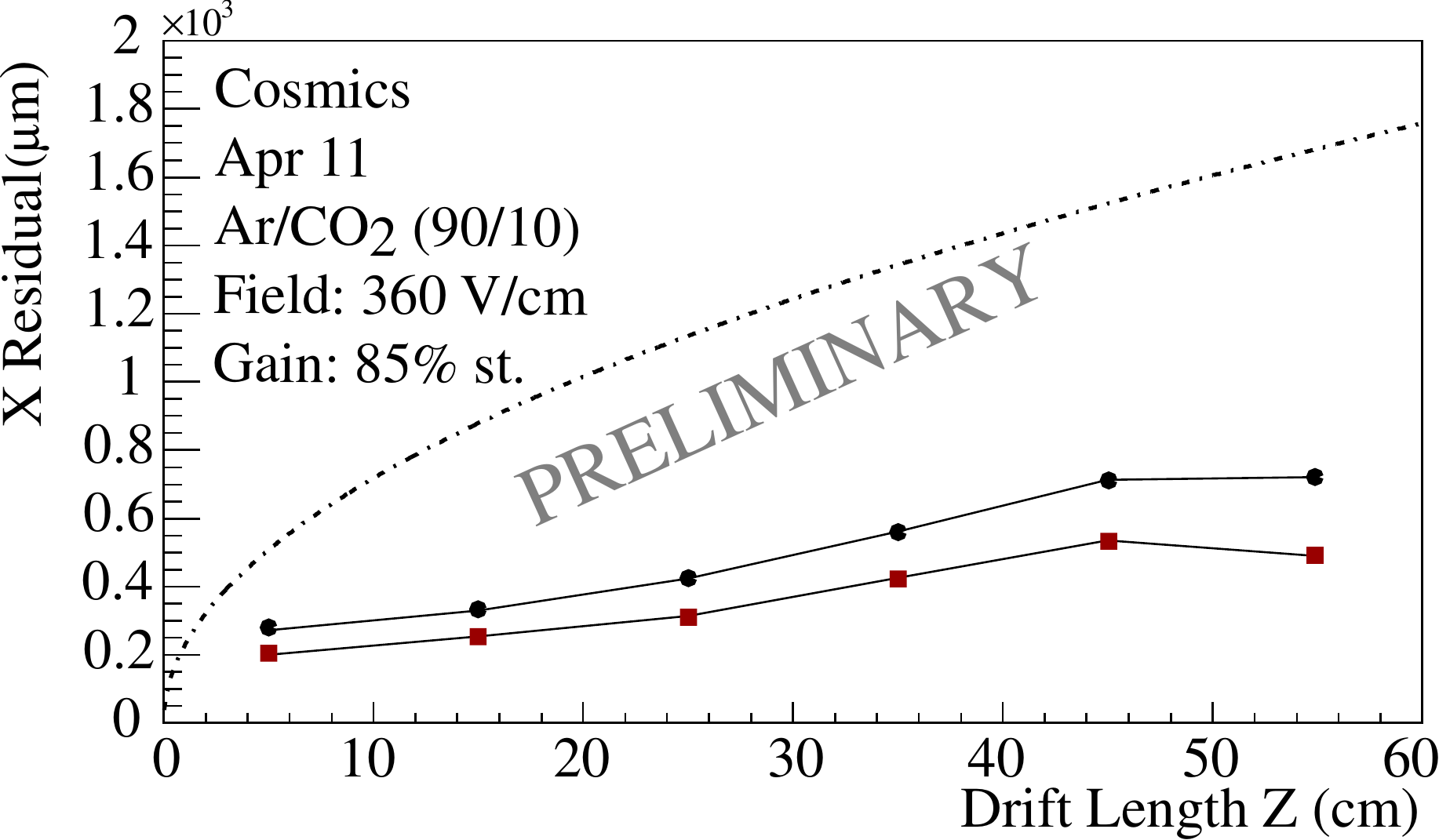}
  \caption{Residual width in $10\,\Cm$ bins of the drift
    length from cosmics tracks. The residual distributions have been
    fitted with a sum of two Gaussians; (red dots) sigma of the narrow
    Gaussian, (black dots) weighted mean of both
    Gaussians. Statistical error bars are included, 
    but smaller than the data points. Systematic errors due to
    clustering and field distortions are not yet included.  
    The dashed line is the transverse
    diffusion for single 
    electrons.}
  \label{fig:fopi.residual}
\end{figure}

\begin{figure}[tbp] 
  \centering 
  \includegraphics[width=0.45\textwidth,keepaspectratio]{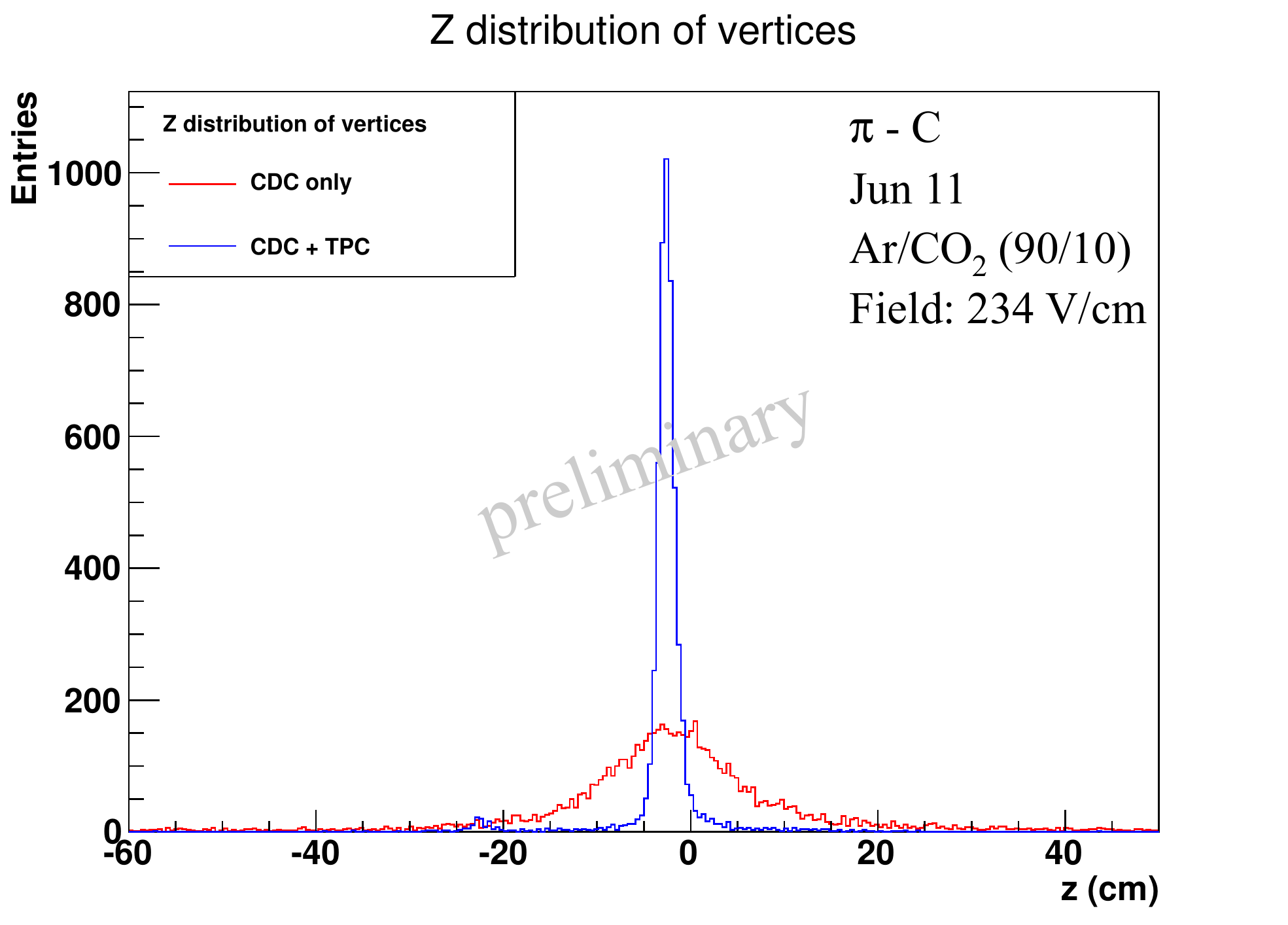}
  \caption{Distribution of reconstructed vertices from $\pi$
    interactions in a $10\,\mm$ thick C target, 
    (red) using the central drift chamber (CDC) only,
    (blue) including the TPC.}
  \label{fig:fopi.vertex}
\end{figure}

In the framework of the PANDA experiment at FAIR a prototype GEM-TPC 
has been built which is the largest detector of this kind to date
\cite{Fabbietti:2010fv,Ball:2012xh}. 
\begin{figure}[tbp] 
  \centering 
  \includegraphics[width=0.5\textwidth,keepaspectratio]{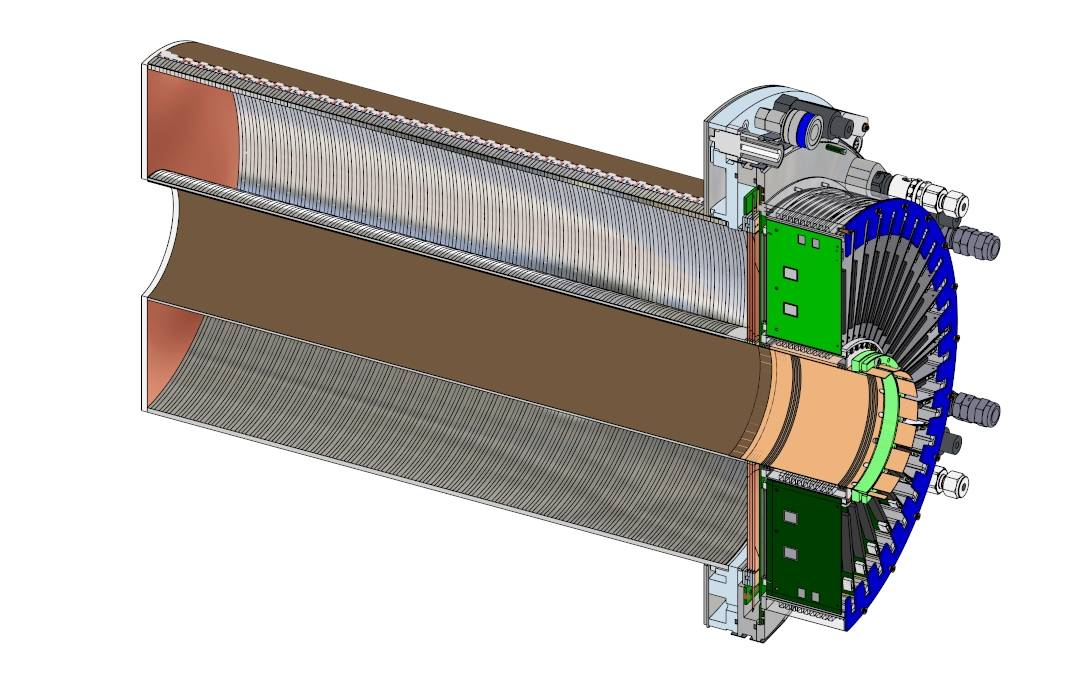}
  \caption{Cross section view of the FOPI GEM-TPC. A uniform electric
    field along the cylinder axis is provided by the cathode plane at
    the left end plane, the inner and outer field cage, and the top
    side of the first GEM at the right and plane.}
  \label{fig:fopi.tpc}
\end{figure}
Figure~\ref{fig:fopi.tpc} shows a cross section of the detector, which 
has a drift length of $725\,\mm$, and an inner (outer)
diameter of $105\,\mm$ ($300\,\mm$). Signals from the triple GEM
amplification region are induced on 10254 hexagonal anode
pads, chosen in order to optimize the spatial resolution for all
possible track directions. The AFTER/T2K chip \cite{Baron:2008zz} is
used to sample the analog signals at a frequency of $16\,\MHz$.  
The detector has 
been commissioned 
with cosmics and particle beams at the FOPI experiment at GSI, and was
recently employed for physics measurements with pion beams. 
The performance of the GEM-TPC, operated with a gas mixture of
Ar/CO$_2$ (90/10), fully matches
the expectations  
with a spatial resolution of $\sim 230\,\upmu\m$ for small drift
distances (Fig.~\ref{fig:fopi.residual}), 
improving the momentum resolution of the existing
spectrometer by $30\%$ and providing more precise vertex coordinates with a
resolution of better than $10\,\mm$ in $z$ direction. 
(Fig.~\ref{fig:fopi.vertex}).  

\begin{figure}[tbp] 
  \centering 
  \includegraphics[width=0.45\textwidth,keepaspectratio]{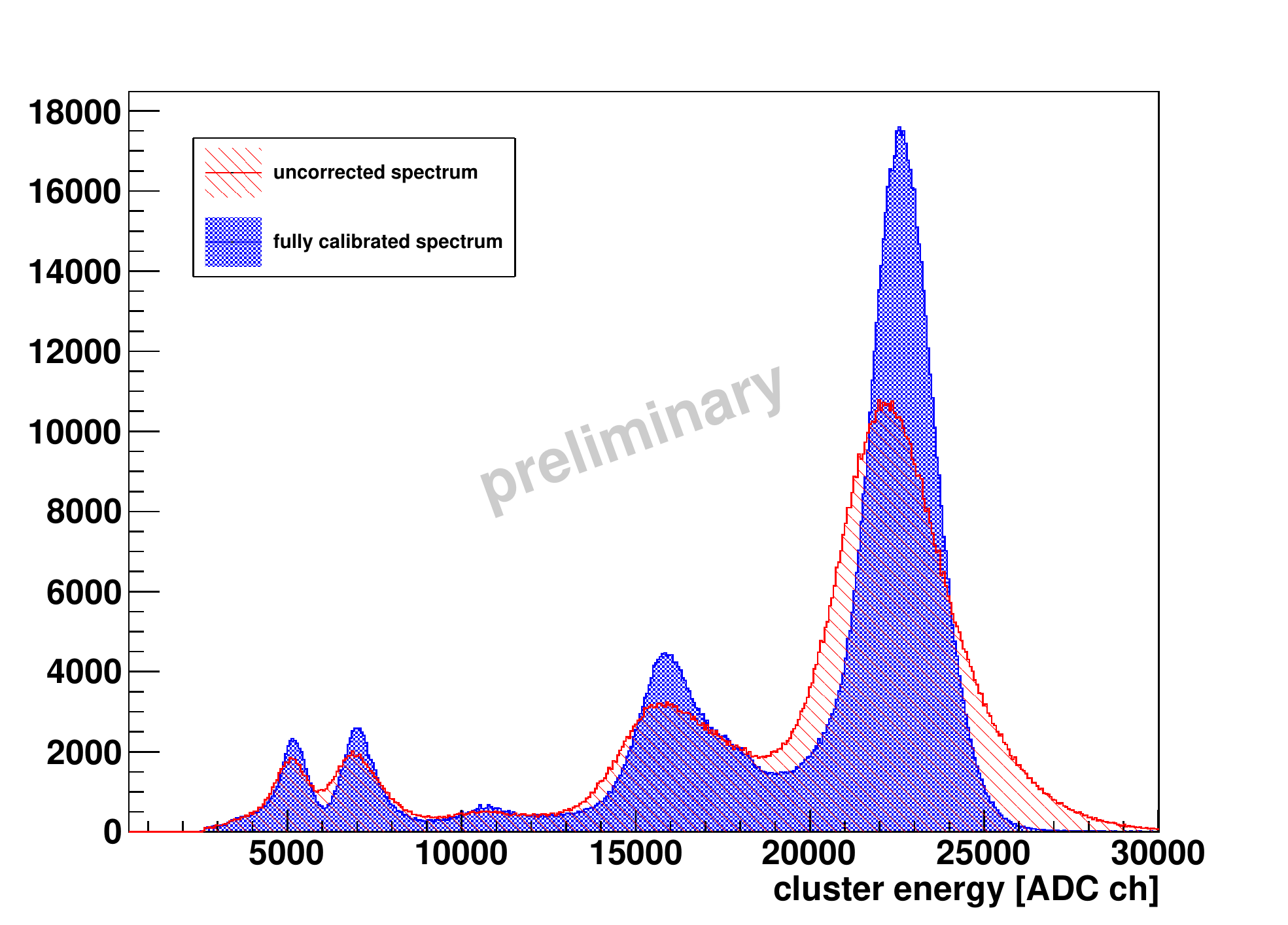}
  \caption{Energy spectrum from decays of a $^{83m}$Kr source
    in the gas volume, obtained with the GEM-TPC in Ar/CO$_2$ (90/10)
    before and after equalization of the gain over 
    the full active area of the GEM.}
  \label{fig:Kr_spectrum}
\end{figure}

\begin{figure}[tbp] 
  \centering 
  \includegraphics[width=0.5\textwidth,keepaspectratio]{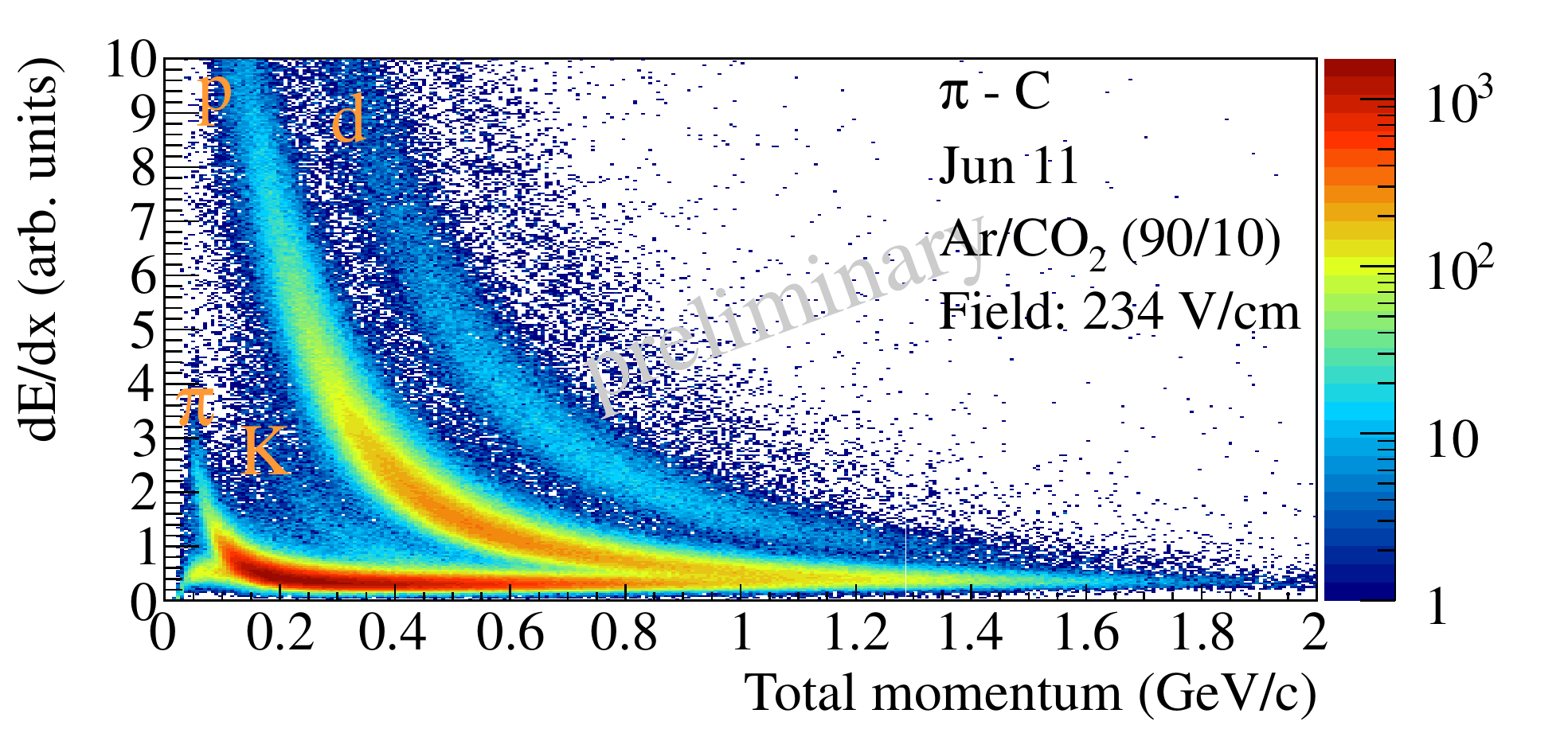}
  \caption{Specific energy loss versus particle momentum measured in
    the GEM-TPC in Ar/CO$_2$ 
    (90/10) in a $0.6\,\T$ magnetic
    field. Here, a truncated mean algorithm cutting away the lowest
    $5\%$ and the highest $25\%$ of sample amplitudes was used.} 
  \label{fig:dEdx_vs_p}
\end{figure}

An important prerequisite for the measurement of the specific energy loss is
the equalization of the gain across the active area of the
detector. As in standard TPCs, this is achieved using decays of a
$^{83\mathrm{m}}$Kr source introduced into the gas 
flow. Figure~\ref{fig:Kr_spectrum} shows the corresponding pulse
height spectrum before and after gain equalization. An energy
resolution of $\Delta E/E=4.4\%$ for the main $41.55\,\keV$ peak is
achieved in Ar/CO$_2$, and similar values are obtained with
a Ne/CO$_2$ mixture. The first measurement of specific energy loss in a
GEM-TPC over a wide momentum range is shown in
Fig.~\ref{fig:dEdx_vs_p}, which has been obtained from data taken with
a pion beam hitting a C target. 
The energy resolution depends on the
number of samples along a track, and reaches $15\%$ for the longest
tracks, in agreement with expectations \cite{Allison:1980vw}.

\section{The ALICE TPC Prototype}
A substantial upgrade of the ALICE detector is foreseen 
until 2018 in
order to fully profit from the increase of luminosity for the
investigation of hot and dense matter using rare probes
\cite{Musa:1475243}. 
The upgrade also includes a replacement of the present gated
MWPC-based amplification system of the TPC
by GEM detectors, and a new continuous untriggered readout of the
detector. 

\begin{figure*}[tbp] 
  \centering 
  \includegraphics[width=0.45\textwidth,keepaspectratio]{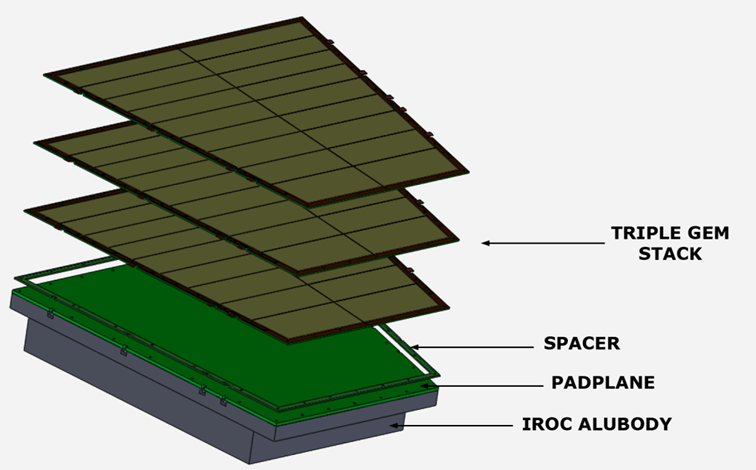}
  \includegraphics[width=0.25\textwidth,keepaspectratio]{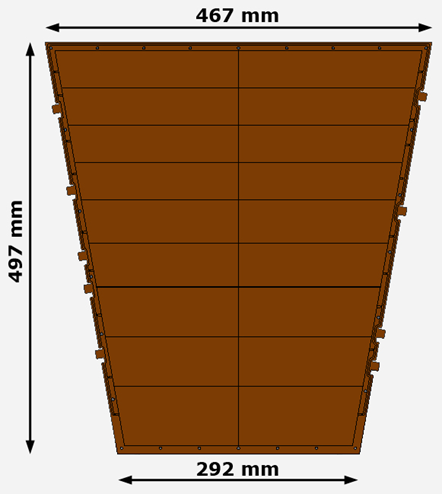}
  \caption{(Left) Exploded view of an ALICE IROC equipped with three
    GEM foils as amplification stage. (Right) Top view of an IROC GEM
    foil, showing the sectorization.}
  \label{fig:alice.iroc}
\end{figure*}

\begin{figure}[tbp] 
  \centering 
  \includegraphics[width=0.45\textwidth,keepaspectratio]{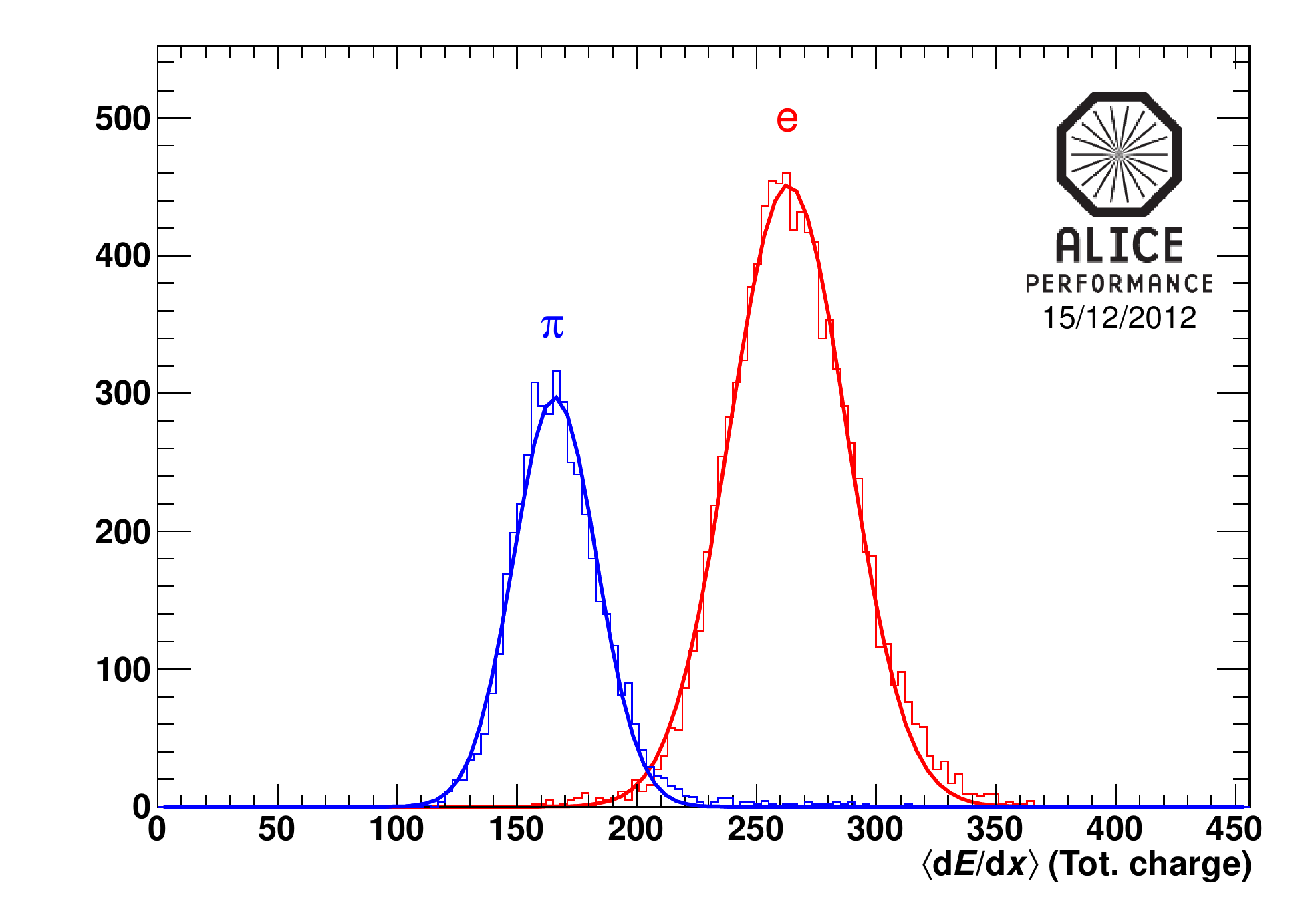}
  \caption{Truncated ($5-70\%$) energy loss distributions for
    electrons and pions 
    with $1\,\GeV/c$ momentum measured with the IROC GEM
    prototype at a gain of $5\cdot 10^{3}$.}
  \label{fig:alice.dEdx}
\end{figure}

\begin{figure}[tbp] 
  \centering 
  \includegraphics[width=0.45\textwidth,keepaspectratio]{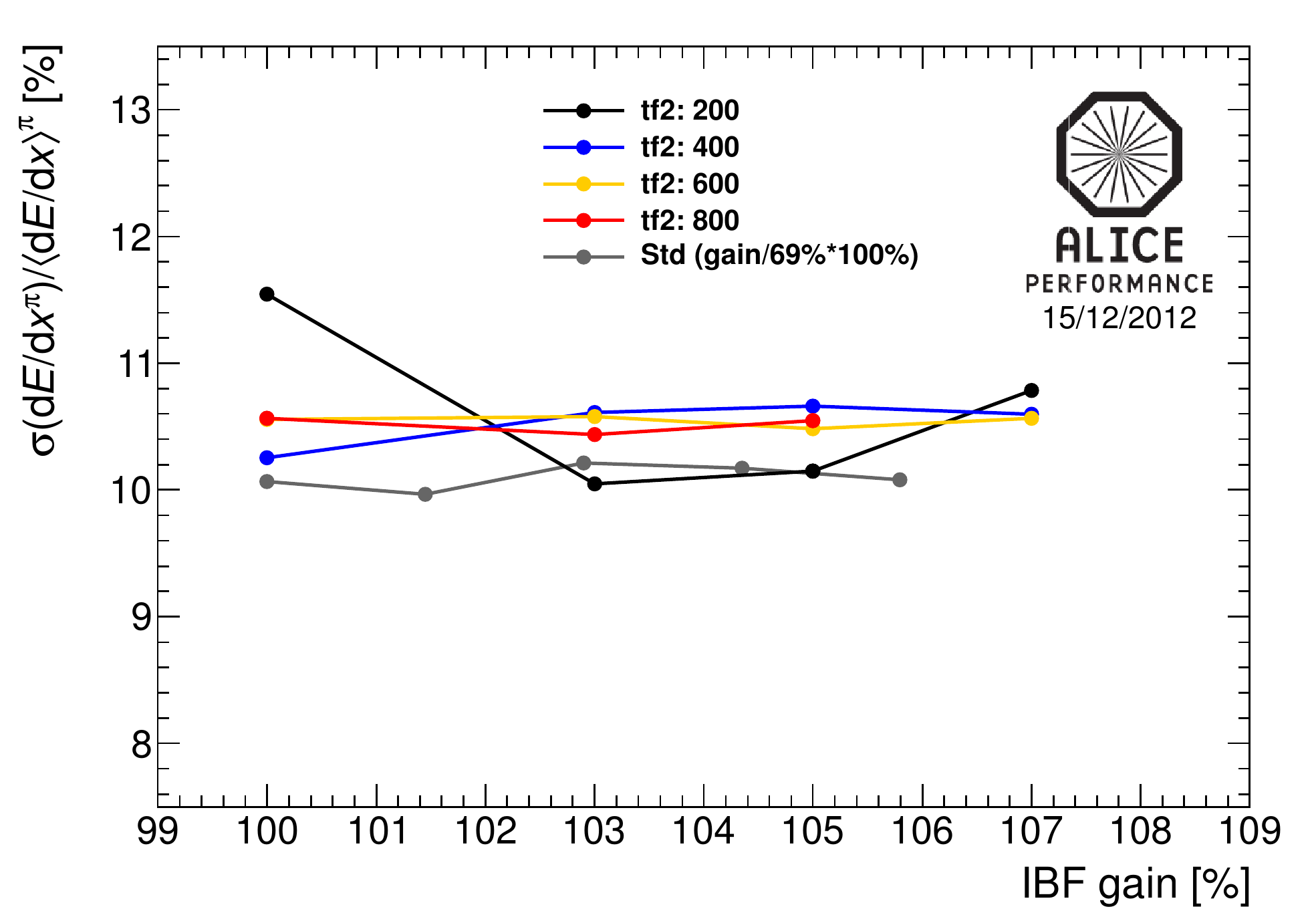}
  \caption{Energy loss resolution for pions with $1\,\GeV/c$ as a
    function of the gain, represented by a scaling factor with respect
    to the settings mentioned in Fig.~\ref{fig:IB-vs-Et2_ne-90_co2-10}.}
  \label{fig:alice.dEdx-vs-scale}
\end{figure}

A first full-size prototype inner readout chamber (IROC) equipped with
GEM foils 
has recently been designed and built (Fig.~\ref{fig:alice.iroc}). 
The GEM foils are of trapezoidal  
shape 
of $504\,\mm$ length and $292\,\mm$ ($467\,\mm$) width of the short
(long) parallel sides, and have been manufactured at CERN using the
newly developed 
single-mask technique \cite{DuartePinto:2009yq}. 
They consist of 18 individually powered sectors
on one side in order to optimize the high-voltage stability of the
chamber. They have been glued on one side onto $2\,\mm$ thick
fiberglass frames with $400\,\mum$ thin spacer 
grids matching the sector boundaries. A new stretching tool making use
of pneumatic frames for a 
homogeneous distribution of forces has been introduced for this purpose. 
The GEM IROC was equipped with three GEM foils with $2\,\mm$ transfer
gaps and an induction gap of $4\,\mm$, matching the design of the FOPI
GEM-TPC. 
For commissioning and testing the detector was placed in a test box
equipped with a drift cathode and a field cage, providing a drift
gap of $11.5\,\Cm$. 
After testing with sources and cosmics, the detector was equipped with 
$1000$ channels of readout electronics based on the PCA16 / ALTRO
chips \cite{EsteveBosch:2003bj,pca16:2009} and  
installed in a beam line at the CERN PS. 
Beam particles ($p$, $\pi^\pm$, $e^\pm$ of momenta between $1$ and
$6\,\GeV/c$) were identified using a Cherenkov and a
lead glass detector. Figure~\ref{fig:alice.dEdx} displays the
energy loss distribution for $1\,\GeV/c$ electrons and negative
pions measured in Ne/CO$_2$ (90/10) with the GEM IROC prototype. A
resolution of  
$10.5\%$ is 
obtained for $1\,\GeV/c$ pion tracks with a 
most probable value of $60$ samples, nearly independent of the gain
between $10^{3}$ and $5\EE{3}$, as can be seen in
Fig.~\ref{fig:alice.dEdx-vs-scale}. 
For $1\,\GeV/c$ electrons, the resolution measured with the prototype
is $9.5\%$ for all gain settings. 
These results are already in very good agreement with the 
value of $9.5\%$ measured for IROCs in the present TPC. 

\section{Conclusions}
A GEM-based TPC opens the possibility of using the powerful
features of such a detector in high-luminosity environments, without
the limits of a gating grid. 
It also presents challenges both to the detector and
to the readout due to the extreme amount of data produced. 
A large GEM-TPC was built and
successfully operated in the FOPI experiment at GSI, fully meeting the
requirements such as improved momentum and vertex resolution, and
yielding
the first measurement of specific energy loss with such a device in a physics
experiment. The ALICE experiment at CERN plans to replace the
MWPC-based readout of the TPC 
with GEM detectors during the LHC shutdown in 2017/2018. A first
prototype IROC was equipped with a triple GEM stack and performed well
during a beam test, confirming the good energy resolution of such a
device. Issues currently being studied in Ne/CO$_2$-based gas mixtures
are the stability of a GEM 
detector in the harsh LHC environment, and the minimization of the ion
backflow. The design of a first prototype GEM OROC (Outer Readout Chamber) is
currently ongoing. 




\providecommand{\href}[2]{#2}\begingroup\raggedright\endgroup

\end{document}